\documentclass[final,authoryear,5p]{elsarticle}

\usepackage{graphicx}

\usepackage{amssymb}

\usepackage{amsmath}
\usepackage{calc}

\usepackage[ps2pdf,%
a4paper=true,%
breaklinks=true,%
colorlinks=true,%
pdfauthor={First Author et al.},%
pdftitle={Template for manuscripts in Advances in Space Research}%
]{hyperref}

\journal{Advances in Space Research}

\begin{document}

\begin{frontmatter}




\title{Ion-atom radiative processes in the solar atmosphere: quiet Sun and sunspots}

\author[label1]{V.A. Sre{\'c}kovi{\' c}}
\ead{vlada@ipb.ac.rs}
\author[label1]{A.A. Mihajlov}
\ead{mihajlov@ipb.ac.rs}
\author[label1]{Lj.M. Ignjatovi{\' c}}
\ead{ljuba@ipb.ac.rs}
\author[label2,label3,label4]{M.S. Dimitrijevi{\'c}}
\ead{mdimitrijevic@aob.rs}
 \address[label1]{University of Belgrade,Institute of Physics, P. O. Box 57, 11001 Belgrade,Serbia}
 \address[label2]{Astronomical Observatory, Volgina 7, 11160 Belgrade 74,Serbia}
 \address[label3]{Observatoire de Paris, 92195 Meudon Cedex, France}
 \address[label4]{IHIS - Technoexperts, Be\v zanijska bb, 11080 Zemun, Serbia}

\begin{abstract}

In the previous works the significance of the symmetric and non-symmetric ion-atom absorption processes in far UV and EUV regions within a model of the quiet Sun atmosphere, has been studied. The considered processes were the processes of the photo-dissociation of the molecular ions H$_{2}^{+}$ and H$X^{+}$ and absorption processes in H+H$^{+}$ and H$+X^{+}$ collisions, where $X$ denotes the metal atom. As the continuation of the previous investigation, these processes are considered also within the corresponding sunspot model. In this work the non-symmetric ion-atom absorption processes are considered with $X=$ Mg, Si, etc. It was analyzed the significance of such processes in far UV and EUV regions in comparison with the concurrent absorption processes, especially with the processes of the photo-ionization of the metal atoms (Na, Mg, Ca, etc.) which were not included in the consideration in the case of the quiet Sun atmosphere. From our analysis it follows that the non-symmetric ion-atom absorption processes considered here, are significant not only for quiet Sun modeling but also for sunspots and should be included {\it ab initio} in both cases.

\end{abstract}


\end{frontmatter}

\parindent=0.5 cm

\section{Introduction}

In previous papers we have studied the influence of some symmetric ion-atom absorption processes in hydrogen case, namely the photo-dissociation of the molecular ion H$_{2}^{+}$ and absorption charge exchange in the H$^{+}$ + H collisions, on the optical characteristics of the quiet Sun atmosphere (see e.q. \citet{mih07a}). Recently, in \citet{mih12,mih13} we have considered this role for the strongly non-symmetric ion-atom absorption processes, also for the quiet Sun atmosphere. The studied processes were absorption charge-exchange and photo-associa\-tion in H$(1s)+X^{+}$ collisions and photo-dissociation of the corresponding H$X^{+}$ molecular ions, where $X$ denotes the metal atom. The influence of processes with $X=$ Mg, Si, Al, etc. was considered and the importance of examined strongly non-symmetric, as well as symmetric and non-symmetric processes together, has been compared with the influence of the known electron-atom (H$^{-}$ and H + free electron) absorption processes. It was demonstrated that the efficiency of the considered ion-atom radiative processes becomes close to the total efficiency of the concurrent electron-ion and electron-atom radiative processes in far UV and EUV regions. It is important that just these spectral regions are very significant because of the daily very strong influence of the corresponding solar radiation on the ionosphere, and by extension the whole of the Earth's atmosphere.

Obtained results in \citet{mih07a,mih12,mih13} demonstrate that in the Solar photosphere the influence of examined processes is very significant and close or larger than the influence of the known electron-atom absorption processes involving H$^{-}$ ion. Moreover, in the quiet Sun photosphere, in far UV and EUV regions, the radiative processes in strongly non-symmetric ion-atom collisions generate the absorption quasi-molecular bands which could significantly influence on the opacity.
The importance of considered processes for the optical characteristics of the quiet Sun atmosphere indicates the necessity to consider their role in the sunspots. Since the temperature is there significantly smaller than in the quiet Sun photosphere, the efficiency of the previously considered symmetric ion-atom radiative processes is much less than
the efficiency of the concurrent electron-atom (H$^{-}$ and H + free electron) absorption processes. For such, umbral Solar plasma conditions, the strongly non-symmetric ion-atom processes with the hydrogen and the relevant metal atoms could have a very important roll. The objective of present investigation is to examine the influence of the above mentioned, strongly non-symmetric processes to the Solar photosphere opacity above sunspots and to compare their influence on the quiet Sun and umbral plasma optical characteristics in far UV and EUV regions.

\section{Quiet Sun}
\label{Sun}

\begin{figure}
\begin{center}
\includegraphics[width=\columnwidth,
height=0.75\columnwidth]{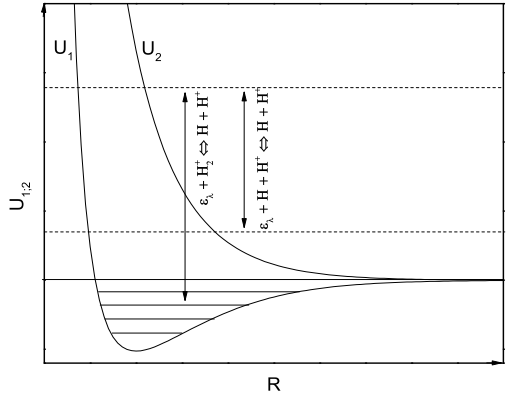} \caption{Schematic presentation of the symmetric processes
caused by the bf- and ff-, radiative transitions}
\label{fig:H2+}
\end{center}
\end{figure}

{\bf The symmetric processes.} In \citet{mih07a} the significant influence of the symmetric ion-atom absorption radiative processes to the solar photosphere opacity in the far UV and EUV regions was demonstrated. Here we keep in mind
the processes of the molecular ion H$_{2}^{+}$ photo-dissociation and absorption charge exchange in the H$^{+}$ + H collisions, namely
\begin{equation}
\label{eq:sim1} \varepsilon_{\lambda}+\textrm{H}_{2}^{+}\longleftrightarrow \textrm{H} + \textrm{H}^{+},
\end{equation}
\begin{equation}
\label{eq:sim2} \varepsilon_{\lambda}+\textrm{H}^{+} + \textrm{H}\longleftrightarrow \textrm{H} + \textrm{H}^{+},
\end{equation}
where H=H(1s), H$_{2}^{+}$ is the hydrogen molecular ion in the ground
electronic state, and $\varepsilon_{\lambda}$ - the energy of a photon with
wavelength $\lambda$. These processes are schematically shown in the figure \ref{fig:H2+}.

As the relevant characteristics of the considered ion-atom absorption processes we will use the
adequately defined spectral absorption coefficients, namely $\kappa^{(bf)}_{sim}(\lambda;h)$ for the photo-dissociation (bound-free) absorption processes (\ref{eq:sim1}), and $\kappa^{(ff)}_{sim}$ $(\lambda;h)$ for the absorption charge-exchange (free-free) processes (\ref{eq:sim2}).
They are defined as the functions of: the wavelength $\lambda$, the height $h$ of the considered layer of the solar photosphere
(with the respect to the referent layer), the local temperature $T$ and the densities of the hydrogen atoms, ions and molecular ions, i.e. $N_{\textrm{H}}$, $N_{\textrm{H}^{+}}$ and $N_{\textrm{H}_{2}^{+}}$
respectively, which are determined within the used model from \citet{ver81}. In the usual way the coefficients $\kappa_{sim}^{(ff)}(\lambda;h)$ and $\kappa_{sim}^{(bf)}(\lambda;h)$  are taken here in the form
\begin{equation}
\label{eq:kapasimff}
\kappa_{sim}^{(ff)}(\lambda;h)=K^{(ff)}_{sim}(\lambda,T)\cdot N_{\textrm{H}}N_{\textrm{H}^{+}},
\end{equation}
\begin{equation}
\label{eq:kapasimbf}
\kappa_{sim}^{(bf)}(\lambda;h)=K^{(bf)}_{sim}(\lambda,T)\cdot N_{\textrm{H}}N_{\textrm{H}^{+}}.
\end{equation}
where $K^{(ff)}_{sim}(\lambda,T)$ and $K^{(bf)}_{sim}(\lambda,T)$ are the corresponding spectral rate coefficients which are determined here as in \citet{mih07a}. Apart of that, the way of the $\kappa_{sim}^{(bf)}(\lambda;h)$ determination is described shortly here.
By definition the spectral absorption coefficient which characterize the bound-free processes of the photo-dissociation of the symmetric molecular ions H$_{2}^{+}$ is given by
\begin{equation}
\kappa_{sim}^{(bf)}(\lambda;h)=\sigma_{sim}^{(phd)}(\lambda,T)\cdot N_{\textrm{H}_{2}^{+}},
\nonumber
\end{equation}
where $\sigma_{sim}^{phd}(\lambda,T)$ is the mean thermal photo-dissociation cross-section
for the molecular ions H$_{2}^{+}$. The photo-dissociation cross-sections $\sigma^{(phd)}_{sim}(\lambda,T)$ are determined here in the same way as in \citet{mih07a}.

Here we will take the coefficient $\kappa^{(bf)}_{sim}(\lambda;h)$ in
an equivalent form, which is more suitable for further considerations, namely
\begin{equation}
\label{eq:kapasimbfA}
\kappa^{(bf)}_{sim}(\lambda;h)=K^{(bf)}_{sim}(\lambda,T)
\cdot N_{\textrm{H}}N_{\textrm{H}^{+}},
\end{equation}
where the spectral rate coefficient $K^{(bf)}_{sim}(\lambda,T)$ is given by
the relations
\begin{equation}
\label{eq:Ksimbf}
K^{(bf)}_{sim}(\lambda,T)=\sigma^{(phd)}_{sim}(\lambda,T)
\cdot \chi_{sim}^{-1}(T),
\end{equation}
\begin{equation}
\label{eq:chisim}
\chi_{sim}(T)=\left [\frac{N_{\textrm{H}}N_{\textrm{H}^{+}}}{N_{\textrm{H}_{2}^{+}}} \right]_{LTE},
\end{equation}

In the last relation the low index "LTE" means that the expression into $[\;]$ is determined under
the local thermodynamical equilibrium (LTE) conditions with given $T$, $N_{\textrm{H}}$ and $N_{\textrm{H}^{+}}$, as it is shown in \citet{mih07a}.
Let us note that these relations are valid also in the case of the used solar photosphere model from \citet{ver81}.


The total contribution of the processes (\ref{eq:sim1}) and (\ref{eq:sim2}) to the solar photosphere opacity is characterized by the total spectral absorption coefficient which is defined by the relations
\begin{equation}
\label{eq:kapasimtot}
\begin{split}
\kappa^{(tot)}_{sim}(\lambda;h)=\kappa^{(bf)}_{sim}(\lambda;h) + \kappa^{(ff)}_{sim}(\lambda;h)=\\
=K^{(tot)}_{sim}(\lambda,T)\cdot N(\textrm{H})N(\textrm{H}^{+}),
\end{split}
\end{equation}
\begin{equation}
\label{eq:Ksimtot}
K^{(tot)}_{sim}(\lambda,T)=K^{(bf)}_{sim}(\lambda,T)+K^{(ff)}_{sim}(\lambda,T),
\end{equation}
where $K^{(bf)}_{sim}(\lambda,T)$ is given by the Eqs. (\ref{eq:Ksimbf}) and (\ref{eq:chisim}).

The efficiencies of these ion-atom absorption processes have to be compared with the efficiencies of the relevant concurrent
processes. Here we keep in mind the processes of the excited hydrogen atom photo-ionization
\begin{equation}
\varepsilon_{\lambda} + \left\{
\begin{array}{l}
\textrm{H}^{*}(n)\\
\textrm{H}^{+} + e^{'}
\end{array}
\right. \to \textrm{H}^{+} + e^{''},
\label{eq:eH+}
\end{equation}
where H$^{*}(n)$ is the hydrogen atom in the one of the  excited states with the principal quantum number $n$, and the especially important absorption processes of the ion H$^{-}$ photo-detachment and the corresponding inverse "bremsstrahlung", namely
\begin{equation}
\varepsilon_{\lambda} + \left\{
\begin{array}{l}
\textrm{H}^{-}\\
\textrm{H} + e^{'}
\end{array}
\right. \to \textrm{H} + e^{''},
\label{eq:eH}
\end{equation}
where H$^{-}$=H$^{-}(1s^2)$ is the hydrogen stable negative ion.
The absorption coefficients which characterize the processes (\ref{eq:eH+}) and (\ref{eq:eH}) are taken here in the form
\begin{equation}
\label{eq:kappaei}
\begin{split}
\kappa_{ei}(\lambda) = K_{ei}(\lambda,T)\cdot N_{e} \cdot N_{\textrm{H}^{+}},\\
K_{ei}(\lambda,T) = \sum_{n\ge 2} \sigma_{n}(\lambda) \frac{N_{\textrm{H}^{*}(n)}}{N_{\textrm{H}}} \cdot \chi_{ei}^{-1} + K_{ei}^{(b)}(\lambda,T)
\end{split}
\end{equation}

\begin{equation}
\label{eq:kappaea}
\begin{split}
\kappa_{ea}(\lambda) = K_{ea}(\lambda,T)\cdot N_{e} \cdot N_{\textrm{H}},\\
K_{ea}(\lambda,T) =\sigma_{phd}^{(-)}(\lambda)\cdot \chi_{ea}^{-1} + K_{ea}^{(b)}(\lambda,T),
\end{split}
\end{equation}

\begin{equation}
\label{eq:chieaei}
\chi_{ea}=\frac{N_{e} N_{\textrm{H}}}{N_{\textrm{H}^{-}}}, \quad \chi_{ei}=\frac{N_{e} N_{\textrm{H}^{+}}}{N_{\textrm{H}}},
\end{equation}
where $\sigma_{phd}^{(-)}(\lambda)$ is the photo-dissociation cross-section of the
negative ion H$^{-}$, $K_{ea}^{(b)}(\lambda,T)$ – the rate coefficient that describes
absorption by (e + H)-collision systems, $\sigma_{n}(\lambda)$ – average
photo-ionization cross-section for the excited atom H$^{*}(n)$,
$K_{ei}^{(b)}(\lambda,T)$ – the rate coefficient that describes absorption by
$(e + \textrm{H}^{+})$-collision systems, and $N_{e}$, $N_{\textrm{H}^{-}}$ and $N_{\textrm{H}^{*}(n)}$ –
respectively the densities of free electrons, negative ions H$^{-}$
and excited atom H$^{*}(n)$. Here, similarly to the case of the molecular ion photo-dissociation,
the coefficients $\chi_{ea}$ and $\chi_{ei}$ are the constants of dissociative-associative and ionization-recombination equilibrium, respectively, for a given $T$.

The relative contributions of the processes (\ref{eq:sim1}) and (\ref{eq:sim2}) together, with
respect to processes (\ref{eq:eH+}) and (\ref{eq:eH}) separately, are characterized here by the quantities
$F^{(sim)}_{ei}$ and $F^{(sim)}_{ea}$ defined by relations
\begin{equation}
\label{eq:Fsimei}
F^{(sim)}_{ei}=\frac{\kappa_{sim}^{(tot)}(\lambda)}{\kappa_{ei}(\lambda)}=\frac{K_{sim}^{(tot)}(\lambda,T)}{K_{ei}(\lambda,T)}\cdot\frac{N_{\textrm{H}}}{N_e},
\end{equation}
\begin{equation}
\label{eq:Fsimea}
F^{(sim)}_{ea}=\frac{\kappa_{sim}^{(tot)}(\lambda)}{\kappa_{ea}(\lambda)}=\frac{K_{sim}^{(tot)}(\lambda,T)}{K_{ea}(\lambda,T)}\cdot\frac{N_{\textrm{H}^{+}}}{N_e},
\end{equation}
where the coefficients $K_{ei}(\lambda,T)$ and $K_{ea}(\lambda,T)$ are given by
Eqs. (\ref{eq:kappaei}) and (\ref{eq:kappaea}). The values of the coefficient $K_{ea}(\lambda,T)$ are determined
by means of data from \citet{sti70} and \citet{wis79} for the emission processes inverse to the absorption
processes (\ref{eq:eH}) and the principle of thermodynamic balance (see \citet{mih11a,mih11b}).
In the case of the absorption processes (\ref{eq:eH+}) the values
of the coefficient $K_{ei}(\lambda,T)$ are determined by exact quantum-mechanical
expressions for the ionization cross-section $\sigma_{n}(\lambda)$
and the quasi-classical expression for the coefficient $K_{ei}^{(b)}(\lambda,T)$
from \citet{sob79}. In the calculations of $\kappa_{ea}(\lambda,T)$ and $\kappa_{ei}(\lambda,T)$
the densities $N_e$, $N_{\textrm{H}}$ and $N_{\textrm{H}^{+}}$, as well as the densities
$N_{\textrm{H}^{*}(n)}$ with $2\le n \le 8$, are taken from \citet{ver81}, while the densities $N_{\textrm{H}^{*}(n>8)}$ were determined by
means of Boltzman formula for given $N_{\textrm{H}^{*}(n = 8)}$ and $T$.

\begin{figure}[ht]
\begin{center}
\includegraphics[width=\columnwidth,
height=0.75\columnwidth]{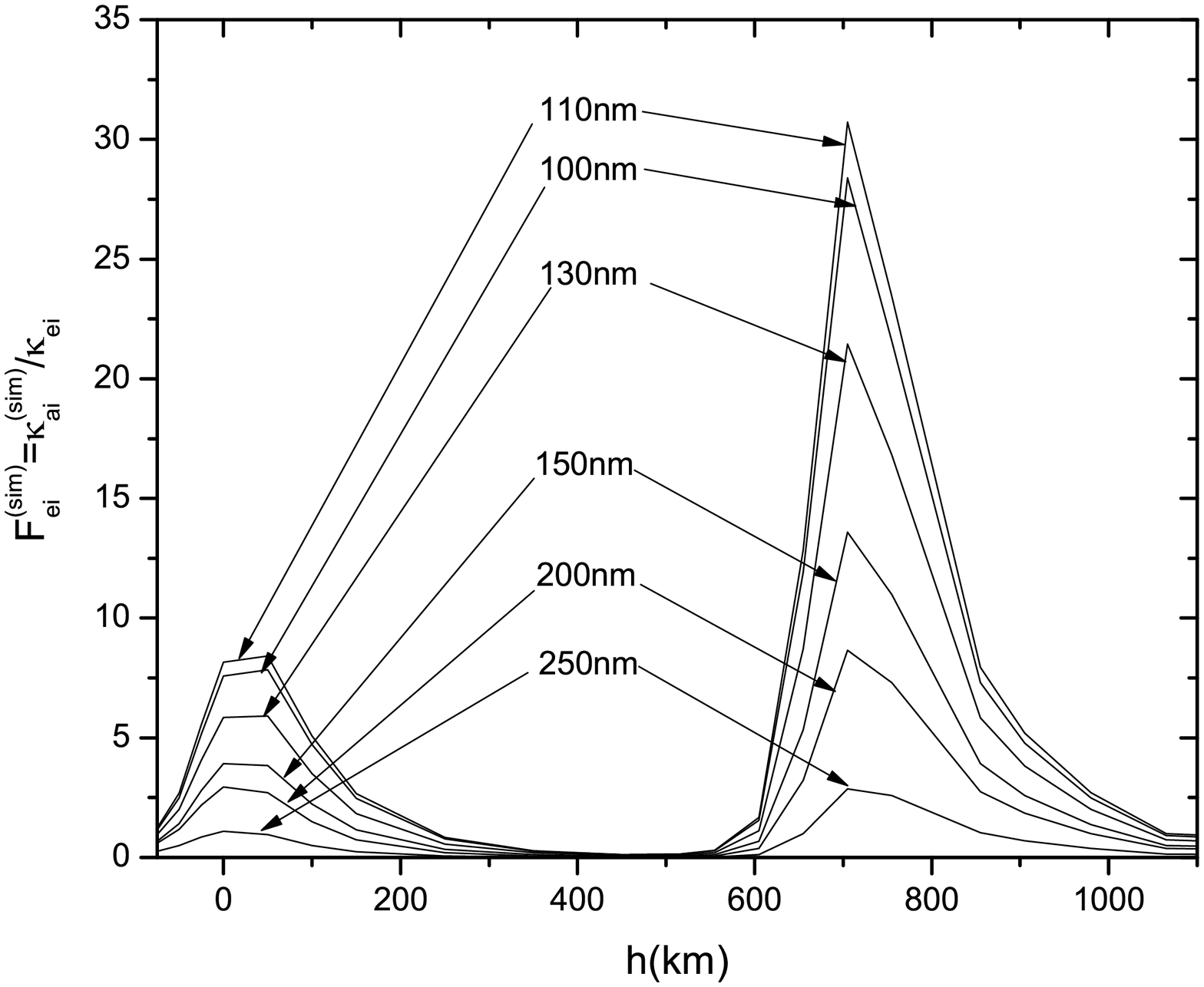} \caption{Behavior of the ratio of the spectral absorption coefficients $F_{ei}^{(sim)}=\kappa_{ia}^{(sim)}/\kappa_{ei}$
as a function of $\lambda$ and $h$.}
\label{fig:Fei}
\end{center}
\end{figure}

\begin{figure}[ht]
\begin{center}
\includegraphics[width=\columnwidth,
height=0.75\columnwidth]{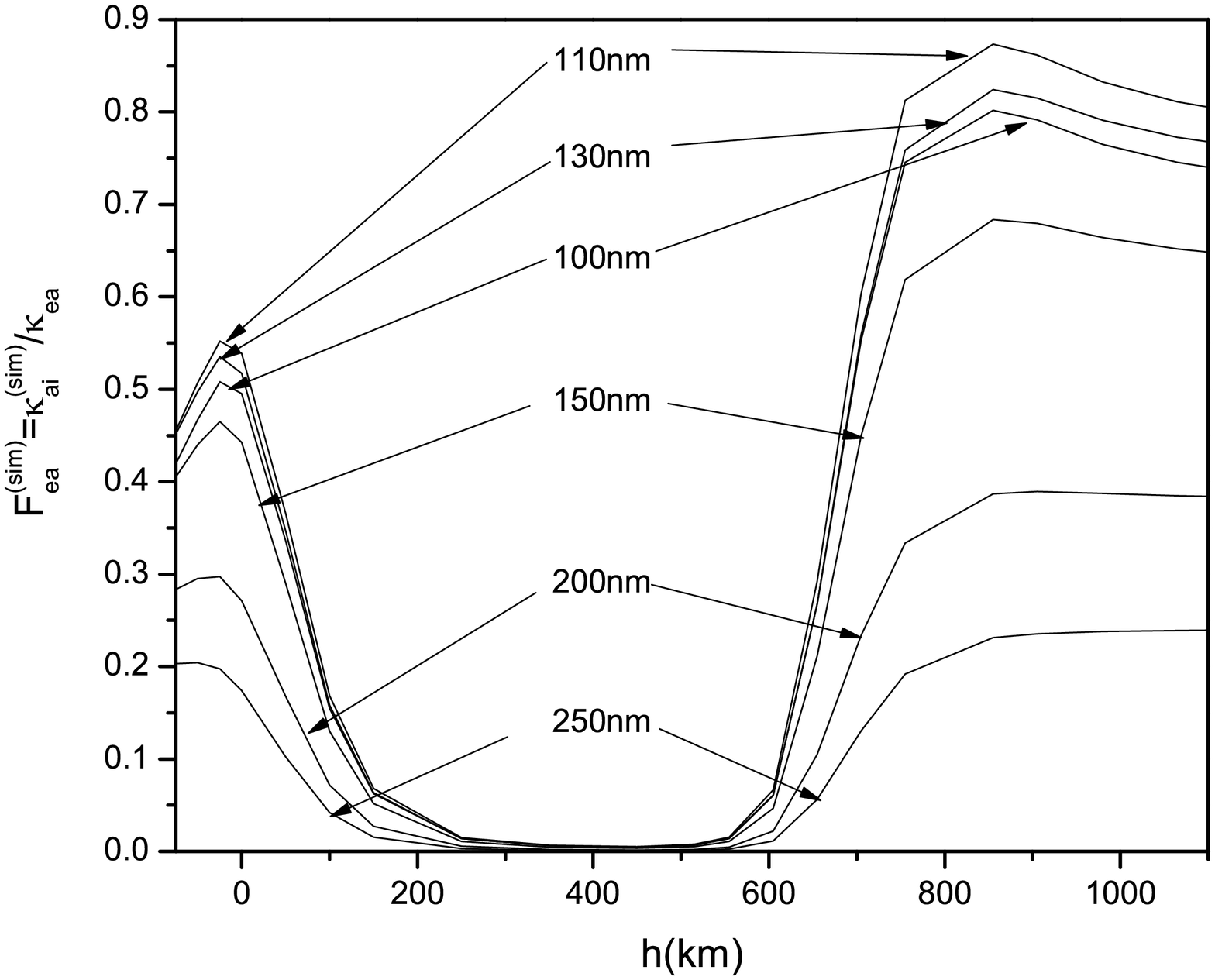} \caption{Behavior of the ratio of the spectral absorption coefficients $F_{ea}^{(sim)}=\kappa_{ia}^{(sim)}/\kappa_{ea}$
as a function of $\lambda$ and $h$.}
\label{fig:Fea}
\end{center}
\end{figure}

The results of the calculations of the ratios $F^{(sim)}_{ei}(h)$ and $F^{(sim)}_{ea}(h)$
for 100 nm $\le \lambda \le$ 250 nm are presented in Figs. (\ref{fig:Fei}) and (\ref{fig:Fea}). The first
of these figures shows that in the significant part of the considered region of altitudes (-75 km $\le h \le$ 1065 km) the absorption processes (\ref{eq:sim1}) and (\ref{eq:sim2}) together dominates with respect to the concurrent electron-ion processes (\ref{eq:eH+}).
This result is the consequence of the fact that in the solar photosphere, in the wide region around the temperature minimum we
have that according to \citet{ver81} the ratio $N_{H}/N_{e} > 10^{3}$.  However Fig. (\ref{fig:Fea}) shows that the absorption processes (\ref{eq:sim1}) and (\ref{eq:sim2}) are very important also with the respect to the main concurrent electron-atom processes (\ref{eq:eH}), since their contribution to the opacity varies from about 10 $\%$ to about 90$\%$ of the contribution of the (\ref{eq:eH}). This means that already the symmetric processes (\ref{eq:sim1}) and (\ref{eq:sim2}) have to be treated as a serious factor that influences the continuous part of the absorption spectra of the Solar atmosphere, particularly in
UV and VUV regions. They influence on the plasma state within particular layers of the Solar atmosphere, which is formed in an interaction with electromagnetic radiation. Due to these reasons the processes (\ref{eq:sim1}) and (\ref{eq:sim2}) are included recently in standard models of the solar atmosphere (see \citet{fon09}).

\begin{figure}
\begin{center}
\includegraphics[width=0.8\columnwidth,
height=0.69\columnwidth]{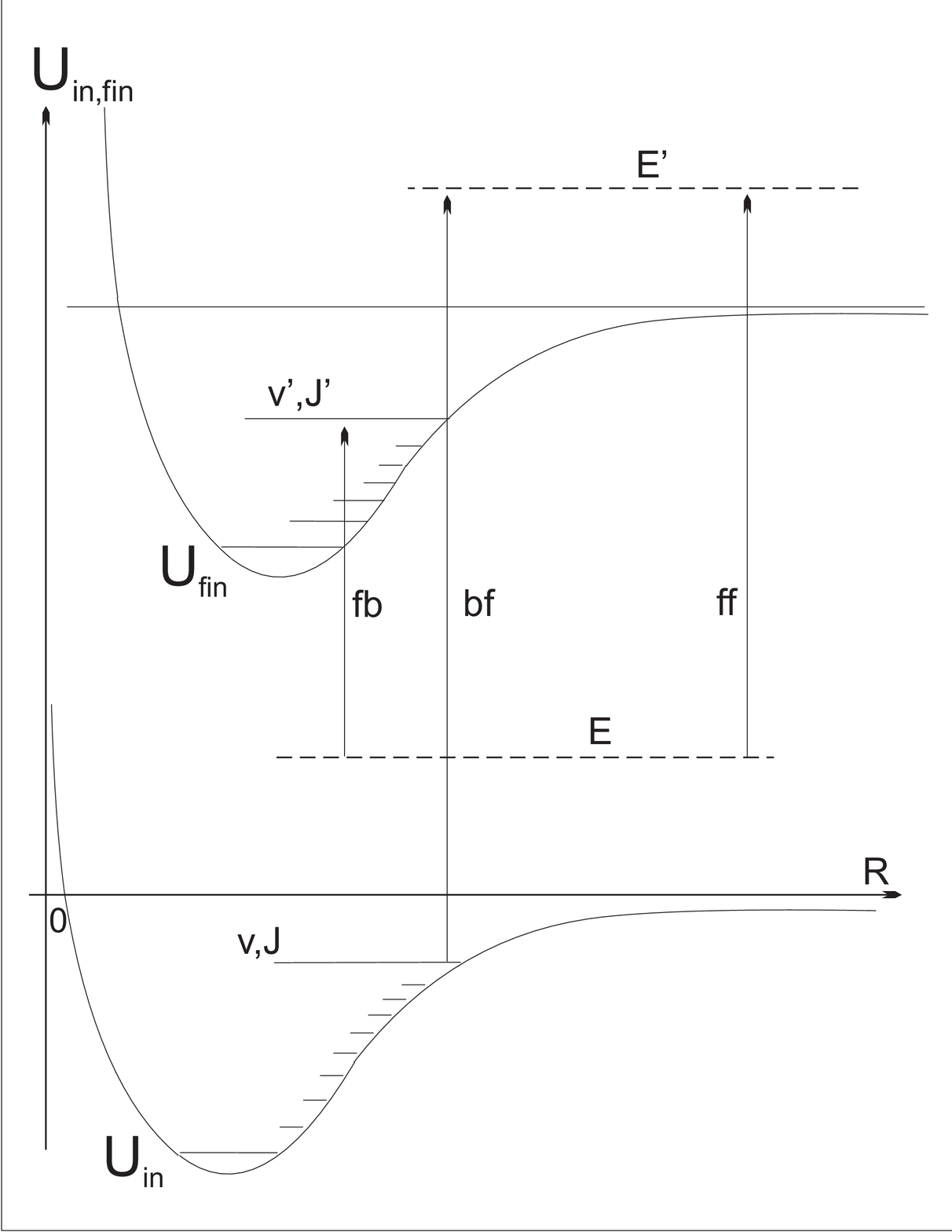} \caption{Schematic presentation of the non-symmetric processes
caused by the bf-, ff-, and fb-radiative transitions}
\label{fig:HX+}
\end{center}
\end{figure}

{\bf The non-symmetric processes.} Among the symmetric ion-atom radiative processes only the processes (\ref{eq:sim1}) and (\ref{eq:sim2}) were taken into account, since the contribution of other such processes to the solar atmosphere opacity could be completely neglected due to the composition of this atmosphere. However, the solar atmosphere contains also the ions $X^{+}$ of several different metals. Because of that in the solar photosphere some of the processes of the photo-dissociation, the absorption charge exchange and the photo-association in the non-symmetric ion-atom collisions, namely
\begin{equation}
\label{eq:nonsim1} \varepsilon_{\lambda}+ \textrm{H}X^{+}\longrightarrow \textrm{H}^{+} + X,
\end{equation}
\begin{equation}
\label{eq:nonsim2} \varepsilon_{\lambda}+ \textrm{H}+X^{+}\longrightarrow \textrm{H}^{+} + X,
\end{equation}
\begin{equation}
\label{eq:nonsim3} \varepsilon_{\lambda}+\textrm{H}+X^{+}\longrightarrow (\textrm{H}X^{+})^{*},
\end{equation}
become possible. Here $X$ and $X^{+}$ are the atom and ion of one of some metals in the ground
states, and H$X^{+}$ and (H$X^{+})^{*}$ are the molecular ions in the electronic
states which are asymptotically correlated with the states of the ion-atom
systems $\textrm{H} +X^{+}$ and $\textrm{H}^{+} + X$ respectively. These processes are schematically shown in Fig.\ref{fig:HX+}

It is clear that the possible partners in these processes are determined by the used solar atmosphere model.
Following \citet{mih12,mih13}, the processes (\ref{eq:nonsim1}) - (\ref{eq:nonsim3}) are considered here within the standard non-LTE (local thermodynamical equilibrium) model C for the quiet solar atmosphere
from \citet{ver81}, since only this model provided  all relevant data (in the
tabular form) needed for the performed calculations. In accordance with chosen
model here, as in \citet{mih13}, in these processes it is taken that $X=$Mg, Si and Al.

Similarly to the symmetric case, we will use the adequately defined spectral absorption coefficients as the relevant characteristics of the considered non-symmetric ion-atom absorption processes, namely: the photo-dissociation (bound-free)
processes (\ref{eq:nonsim1}), absorption charge exchange (free-free) processes (\ref{eq:nonsim2}), and photo-association
(free-bound) processes (\ref{eq:nonsim3}). These processes will be characterized by the basic
spectral absorption coefficients $\kappa^{(bf)}_{\textrm{H}X^{+}}(\lambda;h)$,
$\kappa^{(ff)}_{\textrm{H}X^{+}}$ $(\lambda;h)$ and $\kappa^{(fb)}_{\textrm{H}X^{+}}(\lambda;h)$
respectively. They are defined as the functions of the wavelength $\lambda$ and
the height $h$ of the considered layer of the photosphere,
which is determined (within the used model) by the local temperature $T$ and
local densities $N_{\textrm{H}X^{+}}$, $N_{\textrm{H}}$ and $N_{X^{+}}$ of the molecular ions
H$X^{+}$, hydrogen atoms and ions $X^{+}$, respectively. In the usual way these spectral absorption coefficients are given by relations
\begin{equation}
\label{eq:kapabf}
\kappa^{(bf)}_{\textrm{H}X^{+}}(\lambda;h)=K^{(bf)}_{\textrm{H}X^{+}}(\lambda,T)
\cdot N_{\textrm{H}}N_{X^{+}},
\end{equation}
\begin{equation}
\label{eq:kapaff}
\kappa^{(ff)}_{\textrm{H}X^{+}}(\lambda;h) =
             K^{(ff)}_{\textrm{H}X^{+}}(\lambda,T) \cdot N_{\textrm{H}}N_{X^{+}},
\end{equation}
\begin{equation}
\label{eq:kapafb}
\kappa^{(fb)}_{\textrm{H}X^{+}}(\lambda;h) = K^{(fb)}_{\textrm{H}X^{+}}(\lambda,T)
\cdot N_{\textrm{H}}N_{X^{+}},
\end{equation}
where the spectral absorption rate coefficients $K^{(bf)}_{\textrm{H}X^{+}}(\lambda,T)$, $ K^{(ff)}_{\textrm{H}X^{+}}(\lambda,T)$ and $K^{(fb)}_{\textrm{H}X^{+}}(\lambda,T)$ are determined as in \citet{mih13}.
The spectral absorption coefficient which characterize the bound-free processes of the photo-dissociation of the non-symmetric molecular ions  H$X^{+}$ is given by
\begin{equation}
\kappa_{nsim}^{(bf)}(\lambda;h)=\sigma_{nsim}^{(phd)}(\lambda,T)\cdot N_{HX^{+}},
\nonumber
\end{equation}
where $\sigma_{nsim}^{(phd)}(\lambda,T)$ is the mean thermal photo-dissociation cross-section
for the molecular ions H$X^{+}$. The photo-dissociation cross-sections $\sigma^{(phd)}_{nsim}(\lambda,T)$ is determined here in the same way as in \citet{mih13}.

Here we will take the coefficients $\kappa^{(bf)}_{nsim}(\lambda;h)$ in
an equivalent form, which is more suitable for further considerations, namely
\begin{equation}
\label{eq:kapansimbf}
\kappa^{(bf)}_{nsim}(\lambda;h)=K^{(bf)}_{nsim}(\lambda,T)
\cdot N_{\textrm{H}}N_{\textrm{X}^{+}},
\end{equation}
where the spectral rate coefficient $K^{(bf)}_{nsim}(\lambda,T)$ is given by
the relations
\begin{equation}
\label{eq:Knsimbf}
K^{(bf)}_{nsim}(\lambda,T)=\sigma^{(phd)}_{nsim}(\lambda,T)
\cdot \chi_{nsim}^{-1}(T),
\end{equation}
\begin{equation}
\label{eq:chinsim}
\chi_{nsim}(T)=\left [\frac{N_{\textrm{H}}N_{\textrm{X}^{+}}}{N_{\textrm{HX}^{+}}} \right]_{LTE}.
\end{equation}


From the above mentioned follows that the efficiency of bf-, ff- and fb-absorption channels
together for one of the considered metal species $X$ is characterized by the
partial absorption coefficient $\kappa_{\textrm{H}X^{+}}(\lambda;h)$ which is defined by
the relation
\begin{equation}
\label{eq:kapapar}
\kappa_{\textrm{H}X^{+}}(\lambda;h)=\kappa^{(bf)}_{\textrm{H}X^{+}}(\lambda;h) +
\kappa^{(ff)}_{\textrm{H}X^{+}}(\lambda;h) + \kappa^{(fb)}_{\textrm{H}X^{+}}(\lambda;h).
\end{equation}
Consequently, the total contribution of the mentioned
non-symmetric ion-atom processes to the absorption of the solar radiation on the
height $h$ is described by the total spectral absorption
coefficient which is given by
\begin{equation}
\label{eq:kapansimtot}
\kappa_{nsim}^{(tot)}(\lambda;h) = \sum \kappa_{\textrm{H}X^{+}}(\lambda;h),
\end{equation}
\noindent where the summing is performed over all considered metal species $X$.

Let us note that although in the principle the radiative processes in non-symmetric ion-atom collisions
are known, until recently they were not considered from the aspect of their influences on the optical
characteristics of the solar photosphere. From this reason in \citet{mih13} it was needed to show
that for the solar photosphere the non-symmetric processes (\ref{eq:nonsim1}) - (\ref{eq:nonsim3})
have to bee considered as the important channel of the absorption of the electromagnetic radiation.
For that purpose in \citet{mih13} were performed the calculations of the quantities $F_{ea}^{(sim)}$, given by Eq. (\ref{eq:Fsimea}), and $F^{(tot)}_{ea}(\lambda)$, which is given by the relations
\begin{equation}
\begin{split}
\label{eq:Fsim2}
F^{(tot)}_{ea}(\lambda) = \frac{\kappa_{tot}(\lambda)}{\kappa_{ea}(\lambda)},\\
\kappa_{tot}(\lambda) = \kappa^{(tot)}_{nsim}(\lambda) + \kappa^{(tot)}_{sim}(\lambda),
\end{split}
\end{equation}
where $\kappa^{(tot)}_{sim}(\lambda)$ and $\kappa^{(tot)}_{nsim}(\lambda)$ are defined by Eqs. (\ref{eq:kapasimtot}) and (\ref{eq:kapansimtot}), and $\kappa_{ea}(\lambda)$ - by (\ref{eq:kappaea}) and (\ref{eq:chieaei}). Consequently,
the spectral absorption coefficient $\kappa_{tot}(\lambda)$ characterize the total efficiencies of all, symmetric and non-symmetric, ion-atom absorption processes.
From (\ref{eq:Fsimea}) and (\ref{eq:Fsim2}) one can see that the comparison of the quantities $F_{ea}^{(sim)}$ and
$F^{(tot)}_{ea}(\lambda)$ shows how much the inclusion of the non-symmetric processes increases
in the case of the quiet Sun the total significance of the ion-atom processes in respect to the main concurrent
electron-atom processes (\ref{eq:eH}). It is clear that we will have the similar situation below in Section \ref{sunspot},
in connection of the sunspots. Because of that in Section \ref{sunspot} we will have also to perform the calculations of the
similar quantities.

\begin{figure}[ht]
\begin{center}
\includegraphics[width=\columnwidth,
height=0.75\columnwidth]{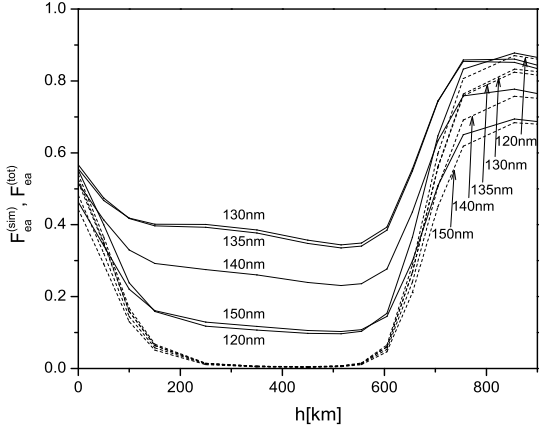} \caption{Quantities $F^{(sim)}_{ea}(\lambda)$
(dashed line) and $F^{(tot)}_{ea}(\lambda)$ (full line), defined in equations (\ref{eq:Fsimea}) and
(\ref{eq:Fsim2}) as the functions of $h$ for the Solar atmosphere for 120 nm $\le \lambda
\le$ 150 nm, \citet{mih13}.}
\label{fig:Fea_tot}
\end{center}
\end{figure}

\begin{figure}[ht]
\begin{center}
\includegraphics[width=\columnwidth,
height=0.75\columnwidth]{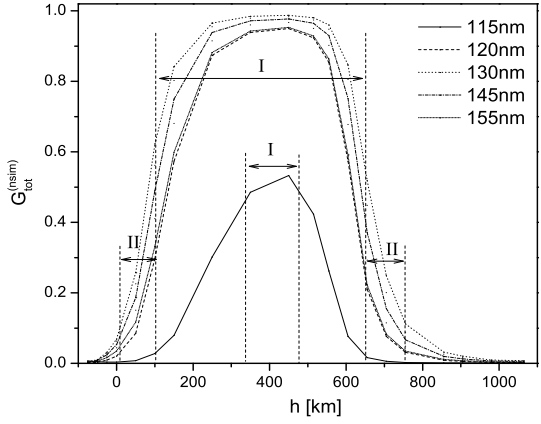} \caption{The presented values of $G^{(nsim)}_{tot}(\lambda)$, given by equation (\ref{eq:Gnsim}), as the function of $h$ for the quiet Sun for 200 nm$ \le \lambda \le$ 230 nm; I and II
are the regions of $h$ where $0.5 \lesssim G^{(nsim)}_{tot}(\lambda)$ and $0.1 \lesssim G^{(nsim)}_{tot}(\lambda) < 0.5$ respectively, \citet{mih13}.}
\label{fig:Gqs}
\end{center}
\end{figure}

The results of the calculations of the quantities $F_{ea}^{(sim)}$ and
$F^{(tot)}_{ea}(\lambda)$ in the case of the quiet Sun are presented in the Fig. \ref{fig:Fea_tot}.
This figure shows that at least in the considered part of the far UV regions the relative contribution to the solar photosphere opacity of the ion-atom radiative processes with the included non-symmetric processes (\ref{eq:nonsim1}) - (\ref{eq:nonsim3}), with respect to the main concurrent electron-atom processes (\ref{eq:eH}), in the neighborhood of the temperature minimum is significantly larger then without these non-symmetric processes. This fact is the consequence of the behavior of the ion component densities (see \citet{ver81}) which is illustrated by figure \ref{fig:Gqs}. In this figure it is presented the behavior of the quantity $G^{(nsim)}_{tot}(\lambda)$ which is given by
\begin{equation}
\label{eq:Gnsim}
G^{(nsim)}_{tot}(\lambda) = \frac{\kappa^{(tot)}_{nsim}(\lambda)}{\kappa_{tot}(\lambda)},
\end{equation}
and describes the relative contribution to the solar photosphere opacity of the non-symmetric processes (\ref{eq:nonsim1}) - (\ref{eq:nonsim3}) with respect to the contribution of all ion-atom absorption processes.

As the result we have that in the case of the quiet Sun all
ion-atom processes, i.e. (\ref{eq:sim1}), (\ref{eq:sim2}) and  (\ref{eq:nonsim1})-(\ref{eq:nonsim3}) together, become now a
serious concurrent to other relevant absorbtion processes in far UV and EUV
regions not only out of the neighborhood of the temperature minimum (see \citet{mih07a}), but within the whole solar photosphere. Of course, these results are significant and for atmospheres of other solar or near solar type stars.

At the end of this section let us note that it was certainly better to perform all needed calculation within some
complete solar photosphere model (which already includes the non-symmetric ion-atom processes), but till now such models
don't exists. Because of that the aim of this work, as well as \citet{mih13}, is to show the necessity of the inclusion
\emph{ab initio} of the processes (\ref{eq:nonsim1}) - (\ref{eq:nonsim3}) in the corresponding models. In this case
the situation is similar to one that existed for a long time in connection with the symmetric ion-atom processes
(\ref{eq:sim1}) and (\ref{eq:sim2}) which were included in the one of the solar atmosphere models only in \citet{fon09}.

\section{Sunspots}
\label{sunspot}
\begin{figure}[h!tbp]
\begin{center}
\includegraphics[width=\columnwidth,
height=0.75\columnwidth]{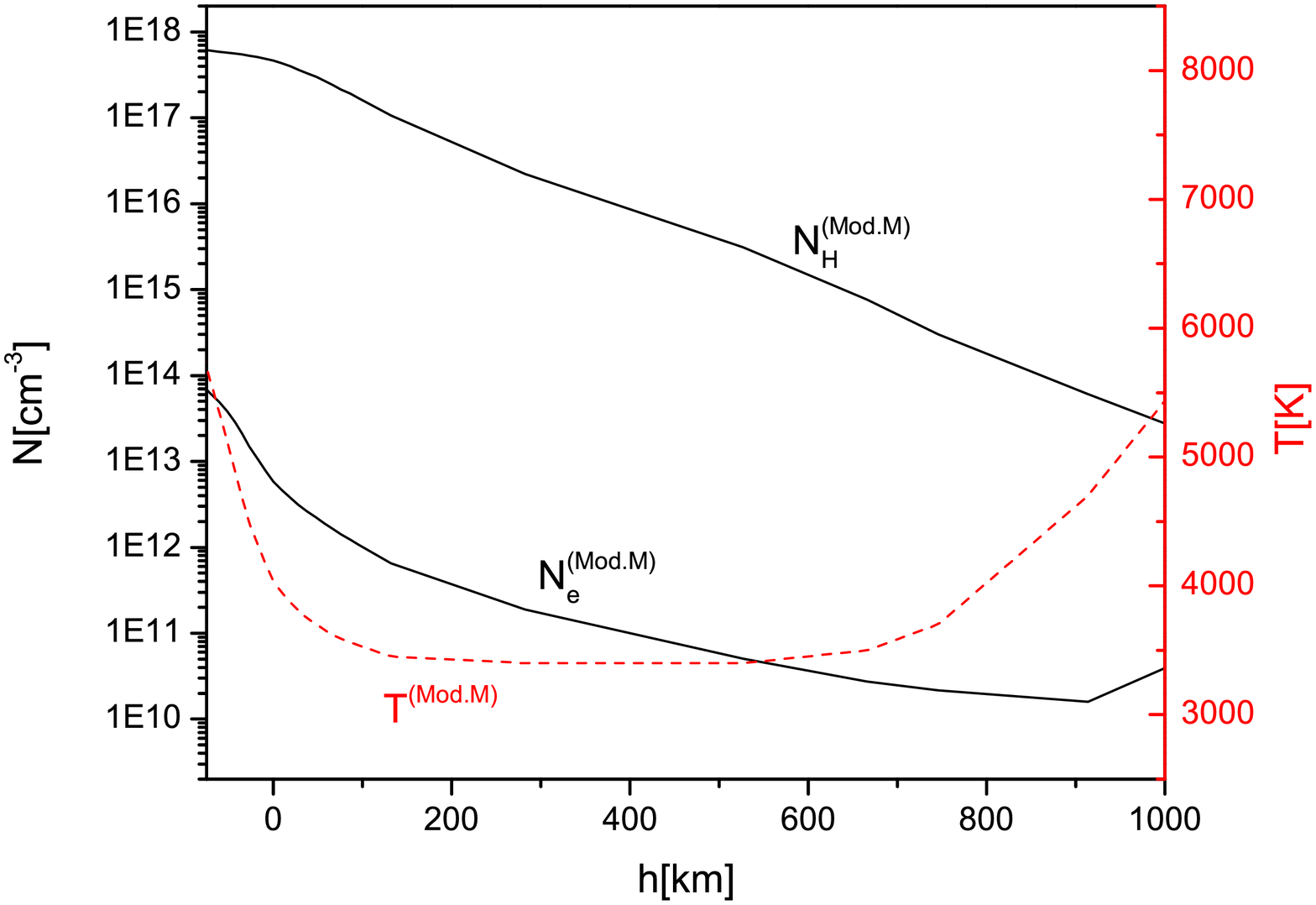} \caption{The behavior of the temperature $T$, $N_{H}$ and $N_{e}$
as a function of height $h$ within the considered part of the solar atmosphere model M of \citet{mal86}.}
\label{fig:Vernaza_vs_Maltby}
\end{center}
\end{figure}

\begin{figure}
\centering
\includegraphics[width=\columnwidth,
height=0.75\columnwidth]{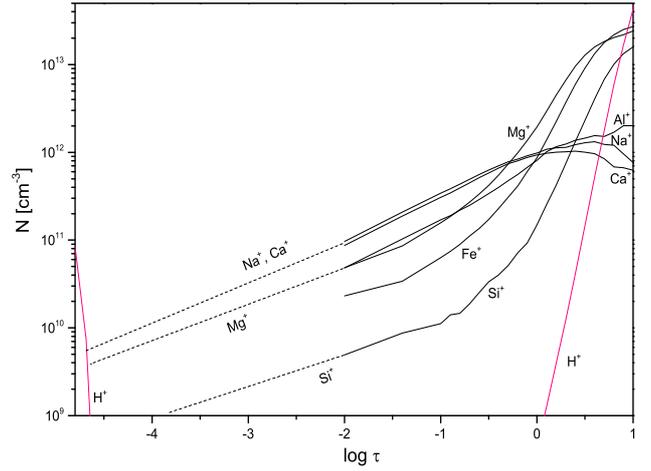}
\caption{The behavior of the densities
$N_{\textrm{H}^+}$ and $N_{X^+}$ of the ions H$^+$ and the metal ions $X^+$ for the
non-LTE model M from \citep{mal86} within the sunspot atmosphere as a function of the Rosseland optical depth log $\tau$.}
\label{fig:Abund_sunspot}
\end{figure}

In this work it is taken into account that the very important parts of the solar atmosphere are its sunspots.
The results above described in connection with the role of symmetric and non-symmetric ion-atom absorption processes in the
atmosphere of the quiet Sun were the main stimulus for the beginning of the investigation of the same processes but within the sunspots.
For that purpose here is used the sunspot model M from \citet{mal86}. This choice is caused, by the fact that by now only this model, among all other models mentioned in the literature, provides all relevant data for the needed calculations. The used model is illustrated by figures \ref{fig:Vernaza_vs_Maltby} and \ref{fig:Abund_sunspot}. In first of these figures the profiles of hydrogen atom and electron densities and the temperature are presented as function of height $h$, while  in the second figure the densities of the hydrogen ion H$^+$ and metal ions $X^+$ are presented as function of Rosseland optical depth log $\tau$ for $X=$ Na, Ca, Mg, Si, Al and Fe.

It is known that in connection with the sunspots it is needed in principle to consider the corresponding magnetic field which is relatively strong in respect to its neighborhood. However, although this magnetic field can influence on many properties of sunspot's plasma, it is in the same time too weak to affect on the kinematics of the ion-atom collisions, at the thermal impact velocities on the very short distances (which are close to ten atomic units), as well as on the interaction of the considered ion-atom systems with the free electromagnetic field. Just because of this fact in this work the presence of a magnetic field is not taken into account.

\begin{figure}
\centering
\includegraphics[height=0.34\textwidth]{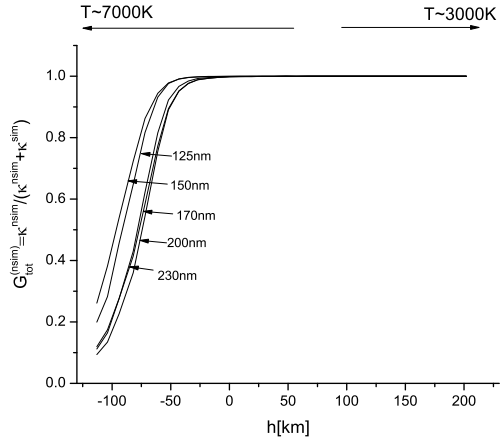}
\caption{The presented values of $G^{(nsim)}_{tot}$, given by equation (\ref{eq:Gnsim}),as the function of
h for the sunspots for $100 \le \lambda \le 230$ nm. }
\label{fig:sim_nsim}
\end{figure}

Within the sunspots the temperature is significantly smaller than in the rest of the solar photosphere. It is clear that because of this fact the relative contribution to the sunspot opacity of the symmetric and non-symmetric processes have to be significantly different with respect to the quiet Sun atmosphere. This is confirmed by the comparison of the Figs. \ref{fig:Gqs} and \ref{fig:sim_nsim} which show the behavior of the quantity $G^{(nsim)}_{tot}$, given by Eq. (\ref{eq:Gnsim}), within the photosphere of the quiet sun and sunspot. Let us note that in the case of the sunspots the total non-symmetric spectral absorption coefficient $\kappa^{(tot)}_{nsim}(\lambda)$ is determined according to Eq.(\ref{eq:kapansimtot}), where $X=$Na, Ca, Mg, Si and Al. The case $X=$ Fe, which is also allowable by the used
model, is not considered here (similarly to \citet{mih13}) because of the absence of the data about the needed characteristics of the molecular ions HFe$^{+}$ and (HFe$^{+})^{*}$.

\begin{figure}
\begin{center}
\includegraphics[width=\columnwidth,
height=0.75\columnwidth]{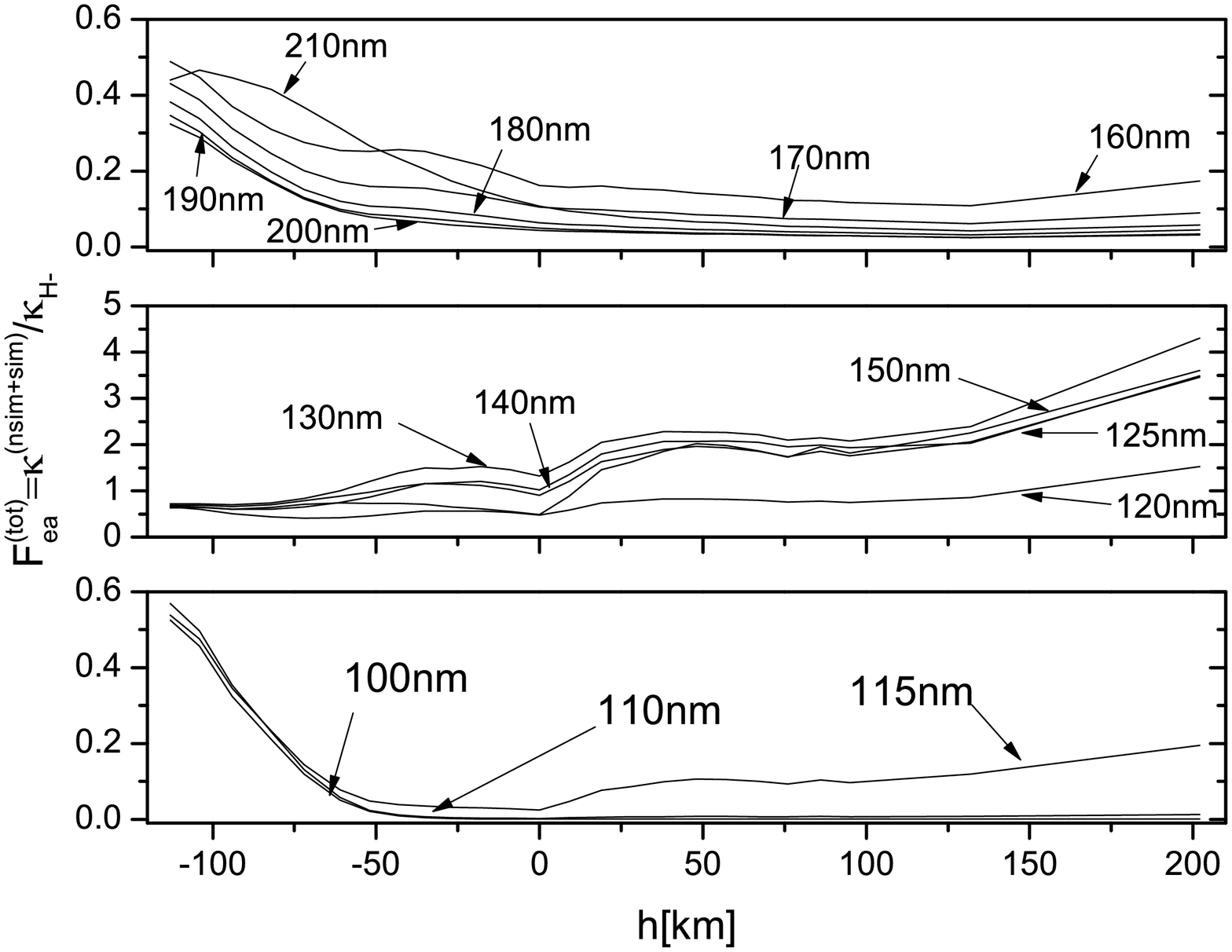}
\caption{Quantity $F^{(tot)}_{ea}(\lambda)$ , defined in equation
(\ref{eq:Fsim2}) as the functions of $h$ for the atmosphere above sunspots for 100 nm $\le \lambda
\le$ 210 nm.} \label{fig:AI_H-b}
\end{center}
\end{figure}

As it was discussed in Section \ref{Sun}, in the case of the sunspot we should also have to perform the calculations of the
quantities $F^{(sim)}_{ea}(\lambda)$ and $F^{(tot)}_{ea}(\lambda)$ given by Eqs. (\ref{eq:Fsimea}) and (\ref{eq:Fsim2}). However,
 within a sunspot the symmetric ion-atom processes (\ref{eq:sim1}) - (\ref{eq:sim2}) are dominant only in the small region of $h$, namely $-100 \textrm{km} \lesssim h \lesssim -75$ km, while in the rest their influence is practically negligible. Because of that it is reasonable to use for umbral plasma the total contribution of ion-atom absorption processes which are characterized by the total ion-atom absorption coefficients $\kappa_{tot}(\lambda)$ defined in Eq. (\ref{eq:Fsim2}). In accordance with this in the case of the sunspots just the total efficiency of the ion-atom absorption processes is compared with the efficiency of the electron-atom absorption processes (\ref{eq:eH}), which were the main concurrent processes in the case of the quiet Sun. For
this reason in the cases of the sunspot we have performed only the calculations of $F^{(tot)}_{ea}(\lambda)$. The behavior of this quantity is presented in Fig.\ref{fig:AI_H-b}. One can see that in the region $h < -50$ km the contribution of ion-atom absorption processes is dominant or at least comparable with respect to the contribution of electron-atom absorption processes (\ref{eq:eH}) in the whole considered spectral region: 100 nm $ \le \lambda \le $ 230 nm. In the rest of a sunspot the ion-atom absorption processes are dominant in respect with the processes (\ref{eq:eH}) in the region 120 nm $ \lesssim \lambda \lesssim $ 150 nm, while for $\lambda <$ 120 nm and $\lambda > 150$ nm the influence of the ion-atom processes is comparable with the influence of the processes (\ref{eq:eH}).

\begin{figure}
\begin{center}
\includegraphics[width=\columnwidth,
height=0.75\columnwidth]{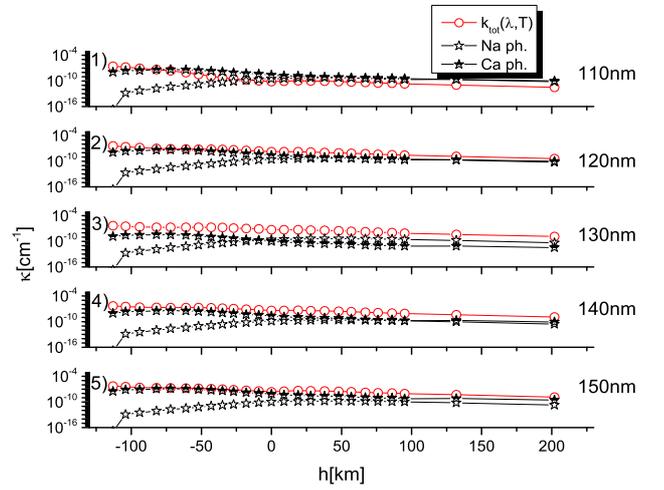}
\caption{Sunspot. Spectral absorption coefficient $\kappa_{tot}(\lambda,T)$, given by
equation (\ref{eq:Gnsim}) and photoionization of Na and Ca atoms for 100 nm$ \le \lambda \le$ 150 nm.}
\label{fig:met}
\end{center}
\end{figure}

The figure \ref{fig:AI_H-b} gives the possibility to estimate the relative significance of the ion-atom absorption processes in the case of a sunspot only with respect to the concurrent electron-atom processes (\ref{eq:eH}). However, according to the model M from \citet{mal86} it is needed in the sunspot case to take into account also several processes of the photo-ionization of the metal atoms ($X=$ Na, Ca, etc.). Consequently, the total contribution (to the  sunspot opacity) of the ion-atom absorption processes has to bee compared with the contribution of the relevant photo-ionization processes of the metal atoms, namely
\begin{equation}
\label{eq:X} \varepsilon_{\lambda}+X \rightarrow e + \textrm{X}^{+},
\end{equation}
where $X=$ Na, Ca, Mg, Si and Al.
In the case of the metal atom photo-ionization processes  (see Eq.(\ref{eq:X})), the spectral absorption coefficients
are denoted with $\kappa_{X}(\lambda,T)$. These coefficients are given by relation
\begin{equation}
\label{eq:kappaX}
\kappa_{X}(\lambda,T) =\sigma_{phi}(\lambda;X)\cdot N_{X},
\end{equation}
where $N_{X}$ is the atom $X$ density and $\sigma_{phi}(\lambda;X)$ is the corresponding photo-ionization cross section.
In connection with this we have two problems. One of them is the absence of reliable data about the molecular ions (HFe$^{+}$, HAl$^{+}$, etc.) while the other problem is the absence of reliable data about the metal atom photo-ionization absorption coefficients. For example, in the case of such known specie as atom Na the data about its absorption coefficients vary in very wide diapason and differ for several times from the experimental data \citep{ise85}. The similar situation exist in the case of other metal atoms. Because of that, within this work, only the relative significance of the considered ion-atom absorption processes and metal atom photo-ionization processes is estimated on few examples illustrated by Figs. \ref{fig:met}.1-\ref{fig:met}.5. These figures show the behavior of the ion-atom total absorption coefficients $\kappa_{tot}(\lambda,T)$ and the corresponding metal atom photo-ionization coefficients $\kappa_{X}(\lambda,T)$ given by Eq.(\ref{eq:kappaX}), with $X=$Na and Ca, in the region 110 nm $\le \lambda \le$ 150 nm. One can see that at least for the considered examples the efficiency of the ion-atom absorption processes is comparable with the efficiency of the considered metal atom photo-ionization processes.

\section{Conclusions}

As first it follows from the presented material that the considered non-symmetric ion-atom absorption processes are significant not only in the case of the photosphere of the quiet Sun but, probably, in the case of sunspots also. This suggest that these non-symmetric ion-atom absorption processes should be included \emph{ab initio} in the solar-photosphere models in both cases: the quiet Sun and sunspots.
As second, one can see that it is necessary a more faster enlargement of the basis of available data concerning characteristics of molecular ions and the coefficients of metal atom photo-ionization, for the sake of further detailed investigations of all relevant absorption processes in sunspots.

\section*{Acknowledgments}
The authors are thankful to the Ministry of Education, Science and Technological Development of the Republic of
Serbia for the support of this work within the projects 176002, III4402.



\clearpage


\end{document}